\documentclass{article}
\pdfoutput=1
\PassOptionsToPackage{square,numbers}{natbib}

\usepackage[final]{neurips_2020}

\usepackage[utf8]{inputenc} 
\usepackage[T1]{fontenc}    
\usepackage{hyperref}       
\usepackage{url}            
\usepackage{booktabs}       
\usepackage{amsfonts}       
\usepackage{nicefrac}       
\usepackage{microtype}      
\usepackage{graphicx}
\usepackage[fleqn]{amsmath}
\usepackage{mathtools}
\usepackage{lipsum}
\usepackage{environ}

\title{Long-Range Seasonal Forecasting of 2m-Temperature with Machine Learning}

\author{%
    Etienne E.~Vos\\
    IBM Research\\
    South Africa\\
    \texttt{etienne.vos\@ibm.com}\\
    \And
    Ashley Gritzman\\
    IBM Research\\
    South Africa\\
    \And
    Sibusisiwe Makhanya\\
    IBM Research\\
    South Africa\\
    \And
    Thabang Mashinini\\
    IBM Research\\
    South Africa\\
    \And
    Campbell Watson\\
    IBM Research\\
    USA}

\begin{document}

\maketitle

\begin{abstract}
A significant challenge in seasonal climate prediction is whether a prediction can beat climatology. We hereby present results from two data-driven models -- a convolutional (CNN) and a recurrent (RNN) neural network -- that predict 2 m temperature out to 52 weeks for six geographically-diverse locations. The motivation for testing the two classes of ML models is to allow the CNN to leverage information related to teleconnections and the RNN to leverage long-term historical temporal signals. The ML models boast improved accuracy of long-range temperature forecasts up to a lead time of 30 weeks for PCC and up 52 weeks for RMSESS, however only for select locations. Further iteration is required to ensure the ML models have value beyond regions where the climatology has a noticeably reduced correlation skill, namely the tropics.
\end{abstract}




\section{Introduction}
Climate change as a result of global warming is a pressing international problem. Concerns are mounting over the significant changes in the variability and extremes of weather, with an increasing possibility of catastrophes from the activation of tipping points in the earth's climate system~\citep{Lenton_etal_2020, IPCC2018}.  There is therefore an increased interest in accurate long-range seasonal forecasts of key climate variables such as surface temperature and precipitation given their relevance to developing strategies that mitigate anticipated seasonal impacts on various sectors, including disaster risk reduction and prevention~\citep{Merryfield_etal_2020}.


Numerical climate models (NCMs) have a long history of being used to produce accurate weather and climate predictions, albeit at the cost of running large and expensive physics-based simulations~(e.g.~\citep{johnson_etal_2019, doi_etal_2016}).  The focus of this work is to investigate how convolutional (CNN) and recurrent (RNN) neural networks can be applied to seasonal forecasting of 2m temperature in lieu of NCMs, and if they are capable of improving upon a generally accepted benchmark that is the $30$-year climatology.  

Previous works \citep{xu_etal_2020, cohen_etal_2019, kamarainen_etal_2019, ham_etal_2019} have shown that these data-driven approaches can perform adequately with respect to physics-based simulations and, in certain cases, surpass them to some extent.  For example, \cite{ham_etal_2019} developed a CNN model with consistently superior all-season correlation skill ($>0.5$) when compared to a state of the art dynamical model (SINTEX-F) \cite{doi_etal_2016} for predicting the Nino3.4 index for lead times of up to 17 months.
\begin{figure}[ht!]\label{fig:figure1}
   \centering
   \includegraphics[width=\linewidth]{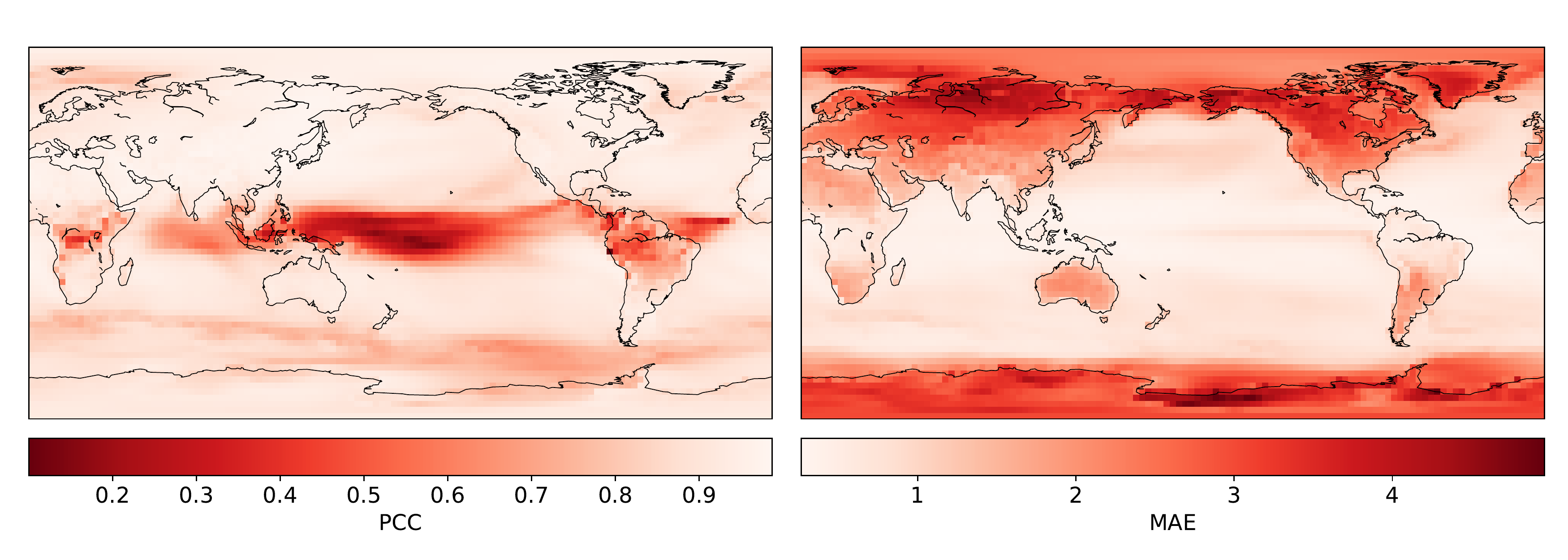}
   \caption{{\it Left:} A global map of the PCC calculated between the ERA5 reanalysis data and the 30-year standard climatology for 2m-temperature.  {\it Right:}  Similar to the left panel, but for MAE. Darker red regions indicate lower PCC or higher MAE values.}
\end{figure}

\section{Methods}
For this work, the ERA5 reanalysis data \cite{hersbach2020era5} is used for training (1979 -- 2007), validation (2008 -- 2011) and testing (2012 -- 2020) of ML models.  Data is pre-processed by regridding global fields of variables from a native spatial resolution of $0.25^\circ\times0.25^\circ$ to $3^\circ\times3^\circ$, as well as aggregating over time from hourly intervals to weekly. The predictor variables used here are $150$mb geopotential ($150$gp), $500$mb geopotential ($500$gp), and $2$m temperature (t2m) fields, the latter of which is also the predictand.

Training and inference for both the CNN- and LSTM-type models were set up in a similar manner:  given a series of inputs  $\textbf{s}_{in}=\{\textbf{x}_{-h_{in}+1}^k, ..., \textbf{x}_0^k \}$ that spans an input-horizon of $h_{in}$ time steps, with $k\in\{\text{t}2\text{m}, 150\text{gp}, 500\text{gp}\}$ and $\textbf{x}$ the global field of variable $k$ at a given time step, the task of the models is to produce predictions $y_{out}$ that estimate the ground-truth values ${x}_{h_f}^{\text{t}2\text{m}}$, which is the 2m temperature for a given target location at a lead time of $h_f$ (forecast-horizon) steps ahead of the latest input time step.  This is done by minimizing the mean squared error (MSE) loss between $y_{out}$ and ${x}_{h_f}^{\text{t}2\text{m}}$ via gradient descent.  The final results are multiple sets of time series predictions of the test data from 2013 up to 2020, each of which is essentially a rolling time-series forecast with a constant $h_f$ lead time, where $h_f\in[1:52]$.  The year 2012 is reserved as a buffer-year for the input horizon.

Predictions are made for single target locations so that separate models had to be trained for all locations.  The following locations were chosen at low and mid/high latitudes across the globe to effectively illustrate the capabilities and limitations of the CNN and LSTM models:

{\setlength{\mathindent}{0pt}
\setlength{\abovedisplayskip}{0pt}
\setlength{\belowdisplayskip}{0pt}
\begin{align*}
    \textbf{Low latitudes: } & \text{Honolulu, USA ($21.3^\circ$N, $157.9^\circ$W); Panama City ($9.0^\circ$N, $79.5^\circ$W)}\\
      & \text{Singapore ($1.4^\circ$N, $103.8^\circ$E), Middle of the Pacific  Ocean ($4.4^\circ$N, $167.7^\circ$W)}\\
    \textbf{Mid/High latitudes: } & \text{Moscow, Russia ($55.8^\circ$N, $37.6^\circ$E); London, UK ($51.5^\circ$N, $0.1^\circ$W)}\\
    & \text{Christchurch, NZ ($43.5^\circ$N, $172.6^\circ$E); Perth, Australia ($32.0^\circ$S,$115.9^\circ$E)}
\end{align*}}

In addition to training a separate model for each location, a separate CNN model was required to make predictions for each lead time.  This setup was mirrored for the LSTM by using a many-to-one model.  The main difference between the CNN and LSTM approaches is that inputs to the CNN are full global fields of all of the predictor variables over an input horizon of $h_{in}=6\,$weeks, whereas inputs to the LSTM are multi-variate time series of the predictor variables extracted at the position of the target location over an input horizon of $h_{in}=52\,$weeks.  

The metrics used to evaluate the final results on the test data are the Pearson Correlation Coefficient (PCC) and the Root Mean Square Error Skill Score (RMSESS), given by the following equations:
\begin{equation}\label{eqn:metrics}
  \text{PCC} = \frac{ \sum_{i=1}^{n}(x_i-\bar{x})(y_i-\bar{y}) }{\sqrt{\sum_{i=1}^{n}(x_i-\bar{x})^2 \sum_{i=1}^{n}(y_i-\bar{y})^2}} \qquad \text{and} \qquad \text{RMSESS} = 1- \frac{\text{RMSE}_{\text{model}}}{\text{RMSE}_{\text{clim}}},\qquad
\end{equation}
where $x$ and $y$ represent the ground-truth and predicted samples, respectively, with $\bar{x}$ and $\bar{y}$ the corresponding sample means over the test data.  The RMSESS compares the model's RMSE to that of the $30$-year climatology.  It is generally difficult to improve upon the climatology in terms of correlation and absolute error.

The CNN architecture consists of 4 convolution blocks (Conv2D$\rightarrow$ReLU$\rightarrow$MaxPool$\rightarrow$Dropout), followed by a 50-unit fully-connected layer and a single-unit output layer.  Fields that comprise $\textbf{s}_{in}$ are stacked as input channels for the CNN. The LSTM architecture consists of an RNN layer with 64 LSTM units, followed by a fully-connected layer with 32 units, and a single-unit output layer. The LSTM does not produce any intermediate predictions, and only produces an output prediction after reading in the full input horizon of 52 weeks.

Anomaly fields with respect to the standard $30$-year climatology were used for all variables during training and inference. The climatology was subsequently added to the outputs to obtain the actual values for final evaluation with the PCC and RMSESS metrics.
\begin{figure}[t!]\label{fig:figure2}
   \centering
   \includegraphics[width=\linewidth]{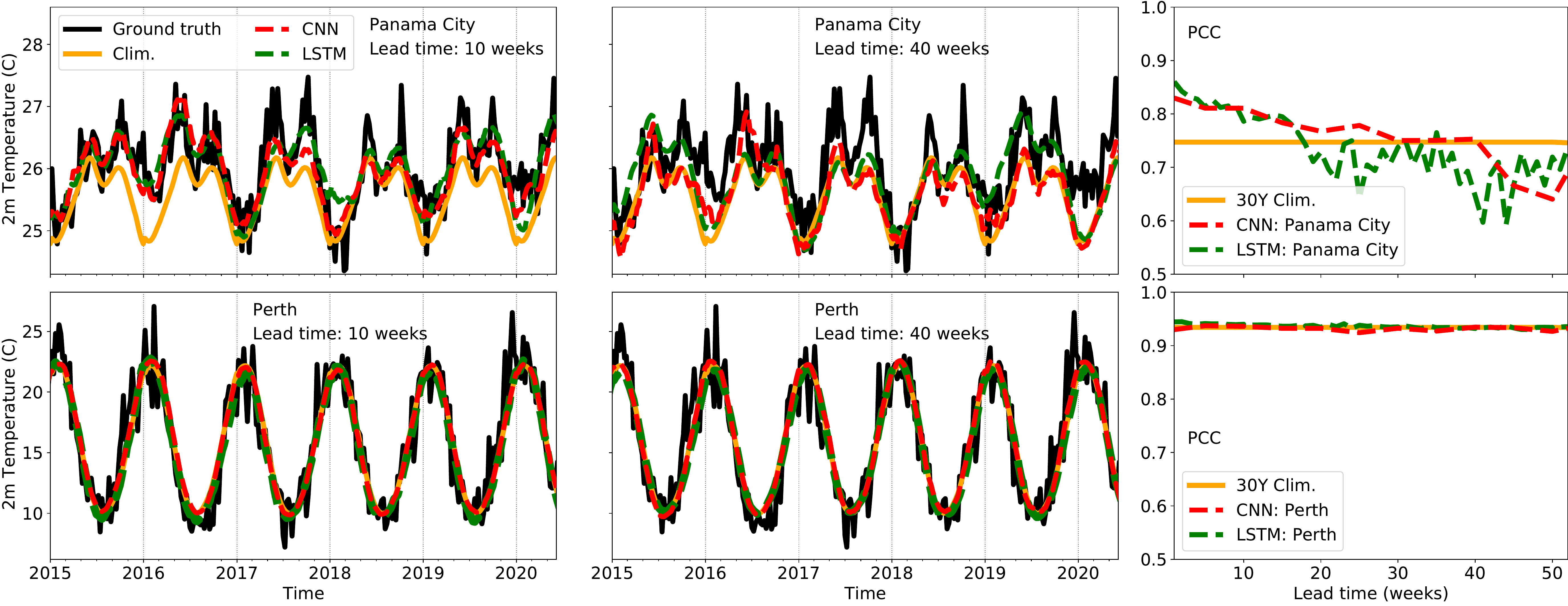}
   \caption{{\it Left \& Center Panels: } Time series plots comparing ERA5 weekly data to the climatology and predictions from the CNN and LSTM for Panama City ({\it top}) and Perth ({\it bottom}).  Left and center panels correspond to predictions at lead times of $10\,$weeks and $40\,$weeks, respectively.  {\it Right Panels:} The PCC at different lead times for Panama City ({\it top}) and Perth ({\it bottom}).}
\end{figure}


\section{Results and Discussion}
The motivation for investigating the two classes of ML models is to allow the CNN to leverage information related to teleconnections in the predictor variables to improve its forecasting skill, while the LSTM should be able to leverage long-term historical temporal information to achieve the same.  In this work, ML results for selected target locations are compared against a baseline prediction, which is the $30$-year standard climatology calculated from weekly-aggregated ERA5 data between 1981 and 2010 (similar to the approach by \cite{janousek_2011}). 

For low latitude locations (near the equator), the climatology has a noticeably reduced correlation skill, as shown in Figure~\ref{fig:figure1}.  Using Panama City as an example, we show in Figure~\ref{fig:figure2} (top panels) that the CNN and LSTM are able to improve on the climatology's PCC skill up to lead times of around $30\,$weeks and $18\,$weeks, respectively.  For a lead time of $10\,$weeks, both models predict the peaks and troughs with reasonable accuracy, capturing to some extent the warmer than usual summers and winters during 2015, 2016 and 2019.  As expected, correlation skill reduces for larger lead times as indicated by the red and green PCC curves that fall below the climatology line.  This can also be seen in the $40\,$week lead time series plot, where CNN and LSTM predictions don't seem to deviate much from climatology, except for a few instances of warmer summers and winters.

In the bottom panels of Figure~\ref{fig:figure2}, Perth is used as an example of a mid/high-latitude location for which the climatology alone already has a PCC skill of $\sim0.93$.  The time series plots show that Perth exhibits a regular annual cycle that is well represented by the climatology so that, for the most part, deviations from the climatology for Perth likely only represent high-frequency noise.  This likely explains why the CNN and LSTM models fail to learn any useful patterns outside of the annual cycle.

\begin{figure}[t!]\label{fig:figure3}
   \centering
   \includegraphics[width=\linewidth]{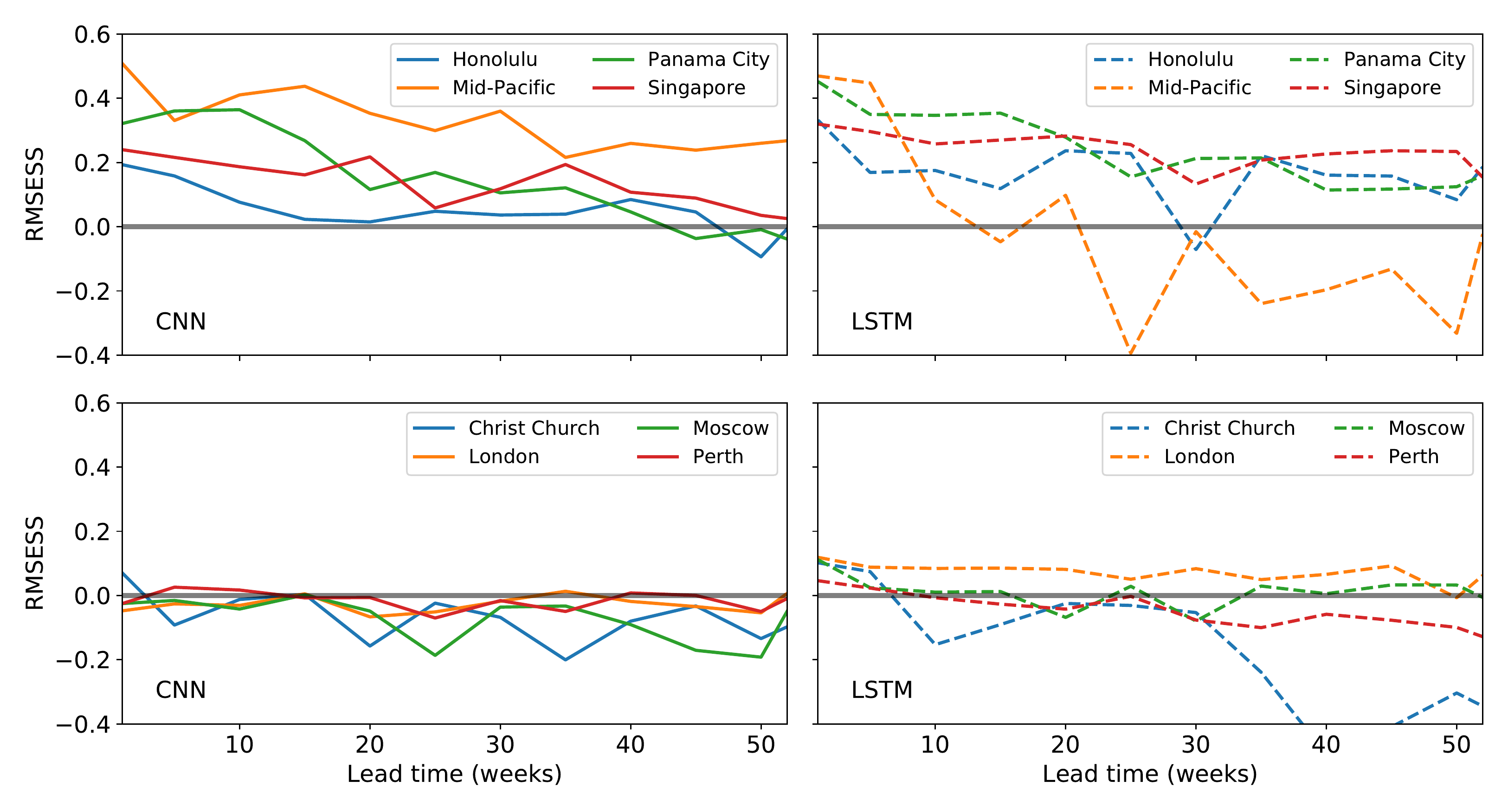}
   \caption{Plots of the RMSESS for lead times of 1 - 52 weeks, and for different locations.  {\it Left Panels:}  RMSESS results from the CNN for locations where predictions have improved skill relative to the climatology ({\it top}), and for locations where predictions have similar or reduced skill than the climatology.  {\it Right Panels:}  Similar to the left panels, but for the LSTM.}
\end{figure}

Figure~\ref{fig:figure3} gives the RMSESS results for the CNN (left panels) and the LSTM (right panels).  These results convey a similar message as those in Figure~\ref{fig:figure2}, but in terms of RMSE.  A RMSESS value $>0$ indicates that the ML model has a lower RMSE than the climatology and, conversely, a value $<0$ means the climatology has a lower RMSE than the model.  For low-latitude locations (top panels) the CNN predictions are able to improve on the climatology for almost all lead times considered.  The same is true for the LSTM, except for the Mid-Pacific location which falls below the climatology for lead times $>20\,$weeks.  Evidently, neither model fares any better than the climatology for the mid/high latitude locations (bottom panels), even at lead times of $<10\,$weeks.  The LSTM does, however, marginally improve on the RMSE for London.

\section{Conclusions and Future Work}
The standard $30$-year climatology, often used as a baseline for seasonal forecasts, does not perform equally well across the globe as highlighted in Figure~\ref{fig:figure1}. However, the $30$-year climatology accurately represents the most important modes of variability for $2$m temperature at all locations with relatively high PCC ($>~0.8$), i.e. outside of the tropics.

Despite the $30$-year climatology being generally difficult to outperform, this study shows that ML methods do achieve comparable, and for some locations, improved, accuracy of long-range temperature forecasts up to a lead time of 30 weeks for PCC and up 52 weeks for RMSESS.  Being able to improve upon such a baseline in the context of seasonal forecasting is an invaluable advantage when considering preparedness against extreme climate events that have characterized climate change impacts over the past two decades.


Other future considerations and improvements on this work include using a more accurate climatology, training on larger datasets like CMIP 5/6, implementing a U-Net approach \cite{ronneberger_etal_2015} in order to generate predictions across the entire globe, as well as to combine the CNN and LSTM models for a unified approach that exploits the spatio-temporal dynamics of the underlying processes.

\begin{ack}
The authors would like to thank Brian White for his mentorship and advice during the preparation of this paper.

\end{ack}



\newpage
\bibliography{./neurips2020_refs.bib}

\begin{thebibliography}{10}

\bibitem{Lenton_etal_2020}
T.~M. Lenton, J.~Rockström, O.~Gaffney, S.~Rahmstorf, K.~Richardson,
  W.~Steffen, and H.~J. Schellnhuber, ``{Climate tipping points — too risky
  to bet against},'' {\em Nature}, vol.~575, pp.~592--595, 2019.

\bibitem{IPCC2018}
{IPCC}, ``{Summary for policymakers},'' in {\em Special Report: Global Warming
  of {$1.5^{\circ}\textrm{C}$}}, p.~32, 2018.

\bibitem{Merryfield_etal_2020}
W.~J. Merryfield, J.~Baehr, L.~Batté, E.~J. Becker, A.~H. Butler, C.~A.~S.
  Coelho, G.~Danabasoglu, P.~A. Dirmeyer, F.~J. Doblas-Reyes, D.~I.~V.
  Domeisen, {\em et~al.}, ``{Current and emerging developments in subseasonal
  to decadal prediction},'' {\em Bulletin of the American Meteorological
  Society}, vol.~101, no.~6, pp.~E869--E896, 2020.

\bibitem{johnson_etal_2019}
S.~J. Johnson, T.~N. Stockdale, L.~Ferranti, M.~A. Balmaseda, F.~Molteni,
  L.~Magnusson, S.~Tietsche, D.~Decremer, A.~Weisheimer, G.~Balsamo, S.~P.~E.
  Keeley, K.~Mogensen, H.~Zuo, and B.~M. Monge-Sanz, ``{SEAS5}: the new {ECMWF}
  seasonal forecast system,'' {\em Geoscientific Model Development}, vol.~12,
  no.~3, pp.~1087--1117, 2019.

\bibitem{doi_etal_2016}
T.~Doi, S.~K. Behera, and T.~Yamagata, ``Improved seasonal prediction using the
  {SINTEX-F2} coupled model,'' {\em Journal of Advances in Modeling Earth
  Systems}, vol.~8, no.~4, pp.~1847--1867, 2016.

\bibitem{xu_etal_2020}
L.~{Xu}, N.~{Chen}, X.~{Zhang}, and Z.~{Chen}, ``{A data-driven multi-model
  ensemble for deterministic and probabilistic precipitation forecasting at
  seasonal scale},'' {\em Climate Dynamics}, vol.~54, no.~7-8, pp.~3355--3374,
  2020.

\bibitem{cohen_etal_2019}
J.~Cohen, D.~Coumou, J.~Hwang, L.~Mackey, P.~Orenstein, S.~Totz, and
  E.~Tziperman, ``{S2S reboot: An argument for greater inclusion of machine
  learning in subseasonal to seasonal forecasts},'' {\em WIREs Climate Change},
  vol.~10, no.~2, p.~e00567, 2019.

\bibitem{kamarainen_etal_2019}
M.~Kämäräinen, P.~Uotila, A.~Y. Karpechko, O.~Hyvärinen, I.~Lehtonen, and
  J.~Räisänen, ``{Statistical learning methods as a basis for skillful
  seasonal temperature forecasts in Europe},'' {\em Journal of Climate},
  vol.~32, no.~17, pp.~5363--5379, 2019.

\bibitem{ham_etal_2019}
Y.-G. Ham, J.-H. Kim, and J.-J. Luo, ``Deep learning for multi-year {ENSO}
  forecasts,'' {\em Nature}, vol.~573, no.~7775, pp.~568--572, 2019.

\bibitem{hersbach2020era5}
H.~Hersbach, B.~Bell, P.~Berrisford, S.~Hirahara, A.~Hor{\'a}nyi,
  J.~Mu{\~n}oz-Sabater, J.~Nicolas, C.~Peubey, R.~Radu, D.~Schepers, {\em
  et~al.}, ``The {ERA5} global reanalysis,'' {\em Quarterly Journal of the
  Royal Meteorological Society}, vol.~146, no.~730, pp.~1999--2049, 2020.

\bibitem{janousek_2011}
M.~Janoušek, ``{ERA}-interim daily climatology,'' {\em ECMWF}, 2011.

\bibitem{ronneberger_etal_2015}
O.~Ronneberger, P.~Fischer, and T.~Brox, ``{U-Net}: Convolutional networks for
  biomedical image segmentation,'' {\em arXiv preprint arXiv:1505.04597}, 2015.

\end{thebibliography}


\end{document}